\documentclass[journal, fleqn]{IEEEtran}
\usepackage{cite}

\ifCLASSINFOpdf
\else
\fi
\usepackage[cmex10]{amsmath}
\usepackage{url}

\usepackage{amsfonts, mathtools, comment}

\newcommand{\sub}[2]{#1_\mathrm{#2}}

\newcommand{\ex}[1]{\mathbb{E}\left[ #1 \right]}
\newcommand{\vb}[1]{\mathbf{#1}}
\newcommand{\mcal}[2]{\mathcal{#1}\left\{ #2 \right\}}
\newcommand{\mr}[1]{\mathrm{#1}}
\newcommand{\norm}[1]{\left\lVert#1\right\rVert}

\def\H{\mathrm{H}}
\def\T{\mathrm{T}}
\def\d{\mathrm{d}}
\DeclarePairedDelimiter\floor{\lfloor}{\rfloor}

\begin{document}
%
\title{Diffuse-field coherence of sensors with arbitrary directional responses}
%
%
%

\author{Archontis~Politis
\thanks{A. Politis is with the Department of Signal Processing and Acoustics, Aalto University, Espoo,
02150 Finland e-mail: archontis.politis@aalto.fi.}
}

\maketitle

\begin{abstract}
Knowledge of the diffuse-field coherence between array sensors is a basic assumption for a wide range of array processing applications. Explicit relations previously existed only for omnidirectional and first-order directional sensors, or a restricted arrangement of differential patterns. We present a closed-form formulation of the theoretical coherence function between arbitrary directionally band-limited sensors for the general cases that a) the responses of the individual sensors are known or estimated, and the coherence needs to be known for an arbitrary arrangement, and b) that no information on the sensor directionality or on array geometry exists, but calibration measurements around the array are available.
\end{abstract}


%
\IEEEpeerreviewmaketitle

\section{Introduction}
%
%
%
%
\IEEEPARstart{T}{he} isotropic diffuse field, a field of uncorrelated plane waves incident with equal probability from any direction and with constant directional power \cite{jacobsen1979diffuse, del2012diffuse}, is a useful simplification of naturally occurring fields with a prominent example being the late reverberant sound in enclosures. It is ubiquitous in a wide variety of acoustical signal processing tasks such as optimal adaptive beamforming, dereverberation and speech enhancement \cite{brandstein2001microphone, mccowan2003microphone, thiergart2014informed}, spatial audio coding and reproduction of array recordings \cite{faller2008microphone, politis2015parametric}, array calibration and unknown geometry inference \cite{mccowan2008microphone}. Such tasks exploit the known structure that the coherence matrix of the diffuse component exhibits between the sensors in the array, in order to either suppress reverberant sound or estimate directional parameters of the sound field \cite{thiergart2012spatial}. 

Contrary to sensor noise, diffuse sound is captured with a degree of coherence between sensors. For omnidirectional elements, the diffuse-field coherence is given by the well-known sinc function of the wavenumber-distance product. For directional elements, the relation becomes more complicated, as it additionally depends on their directional response and orientation, with closed-form solutions available only for first-order gradient sensors \cite{elko2001spatial}.
In practice, omnidirectional sensors can exhibit significant directionality at higher frequencies due to diffraction, depending on the sensor dimensions. Conversely, directional sensors that are assumed to follow some simple nominal first-order directivity, deviate significantly from the assumed response at lower frequencies towards omnidirectional behaviour. In both cases, the assumed diffuse-field coherence matrix of the array becomes inaccurate at the respective frequency ranges, and can introduce estimation errors or performance reduction on the aforementioned applications. Apart from inter-sensor coherence, related cases of interest are the diffuse field-coherence between two beamformers of spaced sub-arrays, e.g. on a hearing aid device, and of head-related transfer functions (HRTFs), suitable for binaural parametric spatial audio rendering \cite{menzer2010stereo}.

This study presents an analytical formulation of the theoretical coherence between sensors or beamformers of arbitrary directionality and orientation, under the practical assumption that their directivity is angularly band-limited. The formulation permits accurate modelling of the coherence matrix of an array, when a model of the directional responses is available, either through an analytical formula, extrapolated by sensor specifications such as polar plots, or measured. The formulation is further demonstrated when there is no knowledge of the sensor responses or the array geometry, but calibration measurements around the array exist. Finally, the case of coherence between beamformers is discussed and an application is given for beam patterns described by the differential form and arbitrarily oriented, for which a closed-form relation previously existed only for the restricted case that the patterns were collinear \cite{elko2001spatial}.

\section{Background}

\subsection{Spherical harmonic transform and coupling coefficients}

The vector of angular spectrum coefficients $\vb{f}$ of a square integrable function $f(\Omega)$ on the unit sphere $S^2$, with $\Omega = (\theta, \phi)$ the inclination and azimuth angle respectively, are given by the spherical harmonic transform (SHT), or spherical Fourier transform, as
\begin{equation}
	\label{eq:sht}
	\vb{f} = \mcal{SHT}{f(\Omega)} =  \int_{\Omega\in S^2} f(\Omega) \vb{y^*}(\Omega) \d\Omega,
\end{equation}
where integration on the unit sphere is denoted as $\int_{\Omega\in S^2} f(\Omega) \d\Omega$ with $\int_{\Omega} \d\Omega = \int_0^{2\pi}\int_0^{\pi}  \sin\theta \d\theta  \d\phi$. The infinite-dimensional basis vector $\vb{y}(\Omega)$, with entries $[\vb{y}(\Omega)]_q = Y_{nm}(\Omega)$ and $q = n^2+n+m+1$, is a vector of complex orthonormalized spherical harmonics (SHs) $Y_{nm}$ of order $n$ and degree $m$, and $[\vb{y^*}(\Omega)]_q = Y^*_{nm}(\Omega)$ denotes its conjugated version. Similarly, the coefficient for a single mode number $(n,m)$ is $[\vb{f}]_q = f_{nm}$. The SHs are defined as
\begin{equation}
	\label{eq:shs}
	Y_{nm}(\Omega) = (-1)^m \sqrt{\frac{(2n+1)}{4\pi} \frac{(n-m)!}{(n+m)!}} P_{nm}(\cos\theta) e^{im\phi}
\end{equation}
with $P_{nm}$ the associated Legendre functions, and $i^2=-1$ the imaginary unit.
The inverse SHT is then
\begin{equation}
	\label{eq:isht}
	f(\Omega) = \mcal{ISHT}{\vb{f}} =  \vb{f}^\T \vb{y}(\Omega).
\end{equation}
The orthonormality of the SHs results in
\begin{equation}
	\label{eq:ortho}
	\int \vb{y}(\Omega) \vb{y}^\H(\Omega)  \d\Omega = \vb{I}
\end{equation}
where $\vb{I}$ is an infinite-dimensional identity matrix.
The Parceval's theorem for the SHT states that
\begin{equation}
	\int f(\Omega) g^*(\Omega)\d\Omega = \vb{g}^\H \, \vb{f} \quad \mathrm{and} \quad \int |f(\Omega)|^2 \d\Omega = \norm{\vb{f}}^2 \label{eq:parceval}.
\end{equation}

In many practical cases order-limited functions are considered, meaning that there is no energy on coefficients above some order $N$, so that $|f_q|^2 = 0$ for $q>(N+1)^2$. We denote the respective $(N+1)^2$-sized coefficient vector as $\vb{f}_N$. For two functions $f(\Omega),g(\Omega)$ band-limited to order $L, M$ respectively, (\ref{eq:parceval}) is limited accordingly by the smaller order of the two. The product $c(\Omega) = f(\Omega)g(\Omega)$ of two such functions is also band-limited to order $L+M$. The spectral coefficients $\vb{c}_{N}$ of such a product function can be computed directly from the coefficients of the member functions $\vb{f}_{L}$ and $\vb{g}_{M}$. More specifically, after applying the SHT (\ref{eq:sht}) on the product and expanding the member functions inside the integral with (\ref{eq:isht}), the $q$--th spectral coefficient of the product is given by
\begin{align}
	\label{eq:gauntproduct}
	c_q &= \int f(\Omega)g(\Omega) Y_q^*(\Omega) \d\Omega \nonumber\\
	& = \vb{f}_L^\T \vb{G}_q \vb{g}_M,  \quad \quad \quad \quad \;\;\, q= 1,2,...,(L+M+1)^2.
\end{align}
The matrix $\vb{G}_q = \int \vb{y}_L(\Omega) \vb{y}^\T_M(\Omega) Y_q^*(\Omega) \d\Omega$ is the $(L+1)^2 \times (M+1)^2)$ matrix that couples linearly the coefficients of the member functions to the coefficients of the product. Its members  $[\vb{G}_q]_{ij} = G_{q'q''}^q$ constitute Gaunt coefficients $G_{q'q''}^q = \int Y_{q'}(\Omega) Y_{q''}(\Omega) Y_q^*(\Omega) \d\Omega$ which can be pre-computed analytically for all mode numbers $(q',q'',q)$ of interest by, e.g., Wigner-3j symbols \cite{sebilleau1998computation}. Example routines to compute Gaunt coefficients and all other operations detailed herein are provided by the author in \cite{politisSHLib}.



Axisymmetric functions are of special practical interest in this work. Such a function can be assumed aligned with the global $z$-axis, in which case $f(\Omega) = f(\theta)$, and is fully described by a reduced set of $N+1$ spectral coefficients
\begin{equation}
	\label{eq:sht_axis}
	\vb{\tilde{f}}_N = \mcal{SHT}{f(\Omega)} =  \int_\Omega f(\Omega) \vb{\tilde{y}}(\Omega) \d\Omega,
\end{equation}
where $[\vb{\tilde{y}}]_{n+1} = Y_{nm |_{m=0}} = \sqrt{(2n+1)/(4\pi)} P_n(\cos\theta)$, $P_n$ is a Legendre polynomial of degree $n$, and $n=0,...,N$. Similarly defined, $[\vb{\tilde{f}}_N]_{n+1} = \tilde{f}_n$.
Based on the SH addition theorem \cite[Eq.14.30.9]{NIST_DLMF}, the coefficients $\vb{f}_N$ of an axisymmetric function oriented at an arbitrary direction $\Omega_0$ are then conveniently given by
\begin{equation}
	\label{eq:steer}
	f_{nm}(\Omega_0) = \sqrt{\frac{4\pi}{2n+1}} \tilde{f}_n \, Y_{nm}^*(\Omega_0).
\end{equation}
Note that all $(N+1)^2$ coefficients are populated in this case.
Finally, another property used later is the relation between the spectral coefficients $\vb{\bar{f}}_N$ of a conjugate spherical function $f^{*} (\Omega)$ and the coefficients of the original function $\vb{f}_N$. Based on the property $Y_{nm}^*(\Omega) = (-1)^m Y_{n(-m)}(\Omega)$ and (\ref{eq:isht}), these can be readily obtained as
 \begin{equation}
	\label{eq:conjSHT}
	\bar{f}_{nm} = (-1)^m f_{n(-m)}^*.
\end{equation}

\subsection{Diffuse field coherence of arbitrary sensors}

An isotropic diffuse field at frequency $\omega$ can be modeled as an amplitude density function $a(\Omega)$ which constitutes a complex random variable with the property
\begin{equation}
\label{eq:diffield}
	\ex{a(\omega, \Omega_1)\, a^*(\omega, \Omega_2)} = \left\{\begin{array}{c}
		\sub{\sigma}{df}^2(\omega),\; \text{for } \Omega_1=\Omega_2\\
		0,\; \text{otherwise}
	\end{array}\right .
\end{equation}
and a power spectral density (PSD) of $\sub{\sigma}{df}^2$ constant with direction. The signal $x(\omega)$ captured by a sensor positioned at $\vb{r}$ with an arbitrary complex directional response $f(\omega, \Omega)$ is
\begin{equation}
\label{eq:diffsig}
	x(\omega, \vb{r}) =  \int f(\omega, \Omega) a(\omega, \Omega) e^{-i\vb{k}^\T(\Omega)\vb{r}} \d\Omega
\end{equation}
where $\vb{k}(\Omega) = -k \vb{n}(\Omega)$, $k=|\omega/c|$ is the wave number, $c$ is the speed of sound and $\vb{n}(\Omega)$ is a unit vector pointing at $\Omega$. The frequency $\omega$ is omitted from the following relations for brevity. The frequency-dependent cross spectral density (CSD) of the signals captured with two sensors with responses $f(\Omega),\, g(\Omega)$ and positioned at $\vb{r}_1,\,\vb{r}_2$ respectively, based on (\ref{eq:diffield}) and (\ref{eq:diffsig}), is
\begin{equation}
\label{eq:diffcsd}
	\Phi_\mr{fg} = \ex{x_1(\vb{r}_1) x_2^*(\vb{r}_2)} = \sub{\sigma}{df}^2 \int f(\Omega)g^*(\Omega)e^{i\vb{k}(\Omega)\vb{r}_{12}}  \d\Omega
\end{equation}
with $\vb{r}_{12} = \vb{r}_{2}-\vb{r}_{1}$. Similarly, the PSD of a single sensor signal, based on (\ref{eq:diffcsd}), is
\begin{equation}
\label{eq:diffpsd} 
	 \Phi_\mr{ff} = \ex{|x|^2} =  \sub{\sigma}{df}^2 \int |f(\Omega)|^2 \d\Omega.
\end{equation}

Finally the diffuse-field coherence between the two sensor signals is given, similarly as in \cite{elko2001spatial}, by
\begin{align}
\label{eq:diffcoh}
	\gamma_\mr{fg} = \frac{\Phi_\mr{fg}}{\sqrt{\Phi_\mr{ff}\Phi_\mr{gg}}} = \frac{\int f(\Omega)g^*(\Omega) e^{i\vb{k}(\Omega)\vb{r}_{12}} \d\Omega}{\sqrt{\int |f(\Omega)|^2 \d\Omega \int |g(\Omega)|^2 \d\Omega}}.
\end{align}

\section{Angular spectrum formulation of diffuse-field coherence}

\subsection{General formulation}
The integral representation of the coherence in (\ref{eq:diffcoh}) can be transformed to a series expansion by means of the SHT. That requires knowledge of the angular spectrum coefficients of the directional responses $\vb{f}, \vb{g}$, and of their respective orientation with regards to the line connecting them, as expressed through the exponential term of (\ref{eq:diffcoh}). Let us assume that the directivities $f(\Omega), g(\Omega)$ are directionally band-limited to some finite-order $N$. The denominator of (\ref{eq:diffcoh}), the product of power spectral densities, is proportional to the angular energies of the directivity functions, expressed through (\ref{eq:parceval}) as
\begin{align}
	\sqrt{\Phi_\mr{ff}\Phi_\mr{gg}} = \sub{\sigma}{df}^2 \sqrt{||\vb{f}_N||^2 ||\vb{g}_N||^2}. \label{eq:sphPSD}
\end{align}
The the cross-spectral density of the numerator can be expressed through (\ref{eq:parceval}) as
\begin{align}
	\Phi_\mr{fg} = \sub{\sigma}{df}^2 \int c(\Omega) b^*(\Omega) \d\Omega = \sub{\sigma}{df}^2 \vb{b}^\H \vb{c} \label{eq:sphCSD}
\end{align}
where $c(\Omega) = f(\Omega)g^*(\Omega)$ is the product of the two directional patterns with angular spectrum $\vb{c}$, and $b(\Omega) =  e^{-i\vb{k}(\Omega)\vb{r}_{12}} = e^{ikd\cos\alpha}$ is the plane wave term with angular coefficients $\vb{b}$. This term is parameterized through the inter-sensor distance $d = ||\vb{r}_{12}||$ and $\alpha$ the angle between the line connecting the two sensors and $\Omega$, $\cos\alpha = \vb{n}^\T(\Omega)\vb{n}(\Omega_r)$ with $\Omega_r = (\theta_r, \phi_r)$ the direction of $\vb{r}_{12}$. 

The spectrum of the product term can be computed from (\ref{eq:gauntproduct}) as
\begin{equation}
	\label{eq:shd_out}
	[\vb{c}]_q = \vb{f}_N^\T \vb{G}_q \vb{\bar{g}}_N,
\end{equation}
where $\vb{\bar{g}}_N$ is the vector of SH coefficients for the conjugate response $g^*(\Omega)$ obtained from (\ref{eq:conjSHT}).
Finally, the angular spectrum of the plane wave is given directly by the plane wave expansion \cite[Eq.10.60.7]{NIST_DLMF},
\begin{equation}
	\label{eq:jacobi}
	e^{ikd\cos\alpha} = \sum_{n=0}^{\infty}\sum_{m=-n}^{m=n} 4\pi i^n j_n(kd) Y_{nm}^*(\Omega_r) Y_{nm}(\Omega)
\end{equation}
where $j_n$ is the spherical Bessel function of order $n$. 
The plane-wave coefficients can be collected in the vector $[\vb{b}]_q = 4\pi i^n j_n(kd) Y_{nm}^*(\Omega_r)$.

Combining (\ref{eq:sphPSD}) and (\ref{eq:sphCSD}), we get the general closed form solution of the diffuse-field coherence of two directional elements
\begin{equation}
	\label{eq:coh_sensorSHT}
	\gamma_\mr{fg}= \frac{ \vb{b}^\H_{2N} \, \vb{c}_{2N} } 
	{\sqrt{\norm{\vb{f}_N}^2 \, \norm{\vb{g}_N}^2}}.
\end{equation}
In case the inter-element distance is small compared to the wavelength ($kd<<1$), so that they can be assumed almost coincident, or when instead of sensors the directional responses describe beamforming patterns that are phase-aligned at the origin of the array, the coherence (\ref{eq:coh_sensorSHT}) simplifies to
\begin{equation}
	\label{eq:cohincidence}
	\gamma_\mr{fg}^\mr{coinc} = \frac{ \vb{g}_N^\H \, \vb{f}_{N} } 
	{\sqrt{\norm{\vb{f}_N}^2 \, \norm{\vb{g}_N}^2}}.
\end{equation}

\subsection{Practical considerations}

For the general relationship of (\ref{eq:coh_sensorSHT}) to have practical significance, the sensor spectra must be of finite dimensions and known at arbitrary orientations. The directional responses of physically realizable sensors can be safely assumed to be angularly band-limited at some order $N$ in their operating frequency range. Knowledge of this order comes from either a band-limited physical model of the sensor response, or by sampling restrictions imposed from a finite number of measurements. In case the sensor response is measured in an anechoic chamber, a minimum of $K>(N+1)^2$ measurements around the sensor should be made in order to capture its spectra of order $N$. In practice, $K$ depends on the measurement grid; usually an equal-angle grid in both azimuth and elevation measuring apparatus is used, in which case the number of points should be in the order of $K\geq 4(N+1)^2$, \cite{driscoll1994computing}. A least-squares approximation to the SHT of (\ref{eq:sht}) can be used on the discrete set of measurements, or for an equiangular grid, the coefficients can be conveniently obtained by a 2-dimensional FFT followed by a unitary transformation \cite{costa2010unified}.

The angular spectra of the individual sensors may be known with regards to their own local frame of reference, appropriate for their specific geometry. To evaluate their coherence in the array coordinate frame, their spectra should be known after their directional responses have been rotated to the desired orientation. If we denote their angular coefficients in their local coordinate frames as $\hat{\vb{f}}_N(\omega), \, \hat{\vb{g}}_N(\omega)$, the rotated spectra can be computed through rotation matrices parameterized by e.g. Euler angles $(\alpha, \beta, \gamma)$, as in 
\begin{equation}
	\vb{f}_N(\alpha, \beta, \gamma) = \vb{T}(\alpha, \beta, \gamma) \hat{\vb{f}}_N.
\end{equation}
For more details on the structure of the rotation matrices, the reader is referred to \cite{driscoll1994computing,rafaely2008spherical}. Apart from the general case, most sensor responses can be assumed axisymmetric on their local frame, for which rotation to a desired direction $\Omega_0$ simplifies considerably and the rotated coefficients $\vb{f}_N( \Omega_0)$ are given by (\ref{eq:steer}). Axisymmetry holds also for many beamformer types, such as ones generated by linear arrays.

\subsection{Measurement-based diffuse-field coherence for unknown arbitrary arrays}

Contrary to the previous cases, in which positions and orientations of sensors with known characteristics may be optimized to achieve certain coherence properties, there may be the case of an array of unknown directional characteristics and with a fixed, and not necessarily known, geometry. It is possible to estimate the diffuse field coherence of an arbitrary array in this case by performing free-field calibration measurements of the array response vector on a dense grid around the array. In this way the true characteristics of the array are modeled appropriately, including arbitrary sensor directionalities and effects of scattering bodies inside the array, which are not captured by simple array specifications. The measured responses can be used directly to estimate the coherence by numerical integration based on (\ref{eq:diffcoh}), but it is advantageous to perform instead the operation in the SH domain, by projecting the measurements to SHs. Both the diffuse-field coherence and an accurate estimate of the interpolated array vector at any direction is obtained this way. Measurement of this angular spectra matrix and its analysis forms the basis of manifold separation techniques for application of direction estimation methods to arbitrary arrays \cite{belloni2007doa, costa2010unified}. Apart from arrays, this case covers also directional scattering functions of interest that are commonly measured, such as HRTFs.

As in the case of a single sensor response, the number of required measurements should be at least $K \geq (N+1)^2$, where $N$ is the maximum order that is appropriate to model the array responses. Since higher-order coefficients decay super-exponentially further away from the centre of the array, a rule-of-thumb for the maximum array order is $N=2\floor{k_\mr{max}R}$, where $k_\mr{max}=2\pi f_\mr{max}/c$ depends on the maximum array operating frequency, and $R$ is the radius of the minimum sphere enclosing all array elements \cite{belloni2007doa}. More advanced approaches can limit further the coefficients, based on information of the sensors' noise power \cite{belloni2007doa}.

The response of the measured arbitrary array with $Q$ sensors to a diffuse field of the the form (\ref{eq:diffield}) is given by 
\begin{equation}
	\label{eq:sensor_out}
	\vb{x}(\omega) =  \int \! a(\omega, \Omega) \vb{d}(\omega, \Omega) \, \d\Omega
\end{equation}
where $\vb{d}(\Omega) = [d_1(\Omega),...,d_Q(\Omega)]^\T$ is the complex response vector of the array to direction $\Omega$. After applying the SHT on the measurement grid, we obtain the matrix of spectral coefficients $\vb{D}_N = [\vb{d}^{(1)}_N, ..., \vb{d}^{(Q)}_N]$, where $\vb{d}^{(q)}_N = \mcal{SHT}{d_q(\Omega)}$. To apply Parseval's theorem (\ref{eq:parceval}) on (\ref{eq:sensor_out}), due to the conjugation of one of the terms, we form the coefficient matrix $\bar{\vb{D}}_N = [\bar{\vb{d}}^{(1)}_N, ..., \bar{\vb{d}}^{(Q)}_N]$ of conjugate responses $d_q^*(\Omega)$ using (\ref{eq:conjSHT}). The array output can be then expressed in the spectral domain as
\begin{equation}
	\label{eq:sensor_outSHT}
	\vb{x} =  \int a(\Omega) \vb{d}(\Omega) \, \d\Omega =  \bar{\vb{D}}_N^\H \vb{a}_N
\end{equation}
Based on (\ref{eq:sensor_outSHT}), the CSD matrix of the array output is given by
\begin{equation}
	\label{eq:sensor_cov}
	\vb{\Phi}_\vb{x} =  \ex{\vb{x}\vb{x}^\H} = \bar{\vb{D}}_N^\H \ex{\vb{a}_N \vb{a}_N^\H}  \bar{\vb{D}}_N.
\end{equation}
Based on the properties of the isotropic diffuse-field of (\ref{eq:diffield}), the covariance matrix of its spectral coefficients up to order $N$ becomes
\begin{align}
	\label{eq:diff_covSHT}
	\vb{R}_\vb{a} &=  \ex{\vb{a}_N\vb{a}_N^\H} = \int\int \ex{a(\Omega)a^*(\Omega')} \vb{y}_N(\Omega) \vb{y}^\H_N(\Omega') \d\Omega' \d\Omega \nonumber\\
	&= \sub{\sigma}{df}^2 \vb{I}_N,
\end{align}
where the orthonormality of the SHs (\ref{eq:ortho}) was used in the derivation. Based on (\ref{eq:diff_covSHT}), the normalized diffuse-field CSD matrix of the array is
\begin{equation}
	\label{eq:sensor_cov2}
	\vb{\Gamma}_\vb{x} = \vb{\Phi}_\vb{x}/ \sub{\sigma}{df}^2 = \bar{\vb{D}}_N^\H \bar{\vb{D}}_N, 
\end{equation}
which results in the diffuse-field coherence between any $(i,j)$ sensor pair in the array
\begin{equation}
	\label{eq:array_coh}
	\gamma_{ij} = \frac{ \floor{\vb{\Gamma}_\vb{x}}_{ij} }{ \sqrt{ \floor{\vb{\Gamma}_\vb{x}}_{ii} \floor{\vb{\Gamma}_\vb{x}}_{jj}} } . 
\end{equation}

\section{Coherence between beamformers and differential patterns}

In many cases it is useful to know the coherence between beamforming patterns, instead of sensors, formed by distributed subarrays, such as e.g. a hearing aid set. To apply the spherical formulation presented above, the angular spectrum of the beamformers should be computed. Recent active research in spherical acoustic processing designs beamformers directly on the spectral domain using spherical arrays, see e.g. \cite{rafaely2015fundamentals}. In this case the spectral coefficients are known and the coherence between beamformers of two such spherical arrays spaced apart is evaluated directly through (\ref{eq:sphPSD}).

Considering uniform linear arrays with symmetrical real weights, a transform that converts beamformer patterns to angular spectra is given in \cite{hafizovic2012transformation}. Following a similar process we demonstrate a transform to obtain spherical spectra for the popular differential arrays, which are able to generate any axisymmetric pattern of up to order $N$ for an array of $N+1$ sensors on their operating frequency range \cite{elko2004differential, de2012design, chen2014design}. The obtainable patterns are described by
\begin{equation}
	\label{eq:differ}
	d(\theta) = \displaystyle\sum_{n=0}^{N} w_{n} \cos^n \theta, \quad\quad\quad \mathrm{with}\quad \sum_{n=0}^{N} w_{n} = 1,
\end{equation}
for a unity maximum response at $\theta=0$. The weights $w_{n}$ determine the pattern and constitute a design choice. Elko in \cite{elko2001spatial} attempted a closed-form solution of the diffuse-field coherence of differential patterns, restricted in the case that their orientation was collinear. However, by transforming the differential weights to angular spectrum coefficients for an axisymmetric function $\tilde{\vb{d}}_N$, the coherence between two such patterns can be computed for any orientation and inter-array distance, based on the results of the previous section.

Relation (\ref{eq:differ}) can be written in vector form as
\begin{equation}
	\label{eq:differ_vec}
	d(\theta) = \vb{w}_N^\T \,\vb{v}_N(\theta)
\end{equation}
with $[\vb{w}_N]_{n+1} = w_n$ and $[\vb{v}_N(\theta)]_{n+1} = \cos^n\theta$. Similarly, the SHT of $d(\Omega)$, assuming it aligned with the $z$-axis results in the $N+1$ angular coefficients $\tilde{\vb{d}}_N$, which according to (\ref{eq:sht_axis})
 \begin{equation}
	\label{eq:sph_vec}
	d(\theta) = \tilde{\vb{d}}_N^\T\, \vb{\tilde{y}}_N(\theta) = \tilde{\vb{d}}_N^\T\, \vb{N}\,\vb{p}_N(\cos\theta)
\end{equation}
where $[\vb{p}_N(\cos\theta)]_{n+1} = P_n(\cos\theta)$ is a basis vector of Legendre polynomials and $[\vb{N}]_{n+1,n+1} = \sqrt{(2n+1)/(4\pi)}$ is a diagonal matrix containing the normalization terms in the definition of the orthonormalized SHs for $m=0$.

To transform the differential weights to the spherical ones, we equate (\ref{eq:differ_vec}) and (\ref{eq:sph_vec}), obtainining
\begin{equation}
	\label{eq:p2c1}
	\tilde{\vb{d}}_N^\T\, \vb{N}\,\vb{p}_N(\cos\theta) = \vb{w}_N^\T \,\vb{v}_N(\theta).
\end{equation}
However, since the Legendre polynomials define a polynomial of degree $n$, $P_n(x) = P_{0n} + P_{1n}x + ... + P_{nn}x^n$, we can define the matrix of Legendre polynomial coefficients $\vb{P}_N$ of size $(N+1)\times(N+1)$, where its columns are the polynomial coefficients of succeeding degrees. Explicit formulas for the polynomial coefficients can be found, e.g., in \cite[Eq.22.3.8]{abramowitz1972handbook}.
The vector $\vb{p}_N$ can then be expressed as
\begin{equation}
	\label{eq:p2c2}
	\vb{p}_N(\cos\theta) = \vb{P}_N^\T \,\vb{c}_N(\theta)
\end{equation}
which, by replacing (\ref{eq:p2c2}) in (\ref{eq:p2c1}), gives the angular spectrum coefficients of the differential patterns with respect to the design weights
\begin{equation}
	\tilde{\vb{d}}_N  = \vb{N}^{-1}\, \vb{P}_N^{-1}\, \vb{w}_N,
\end{equation}
with $[\vb{N}^{-1}]_{n+1,n+1} = \sqrt{4\pi/(2n+1)}$. After the (unrotated) spectral coefficients have been found, coefficients for an arbitrary orientation of the differential pattern to any direction $\Omega_0$ can be derived by using (\ref{eq:steer}). Finally, the diffuse field coherence of two such patterns can be further found by following the procedure outlined in (\ref{eq:sphPSD}--\ref{eq:coh_sensorSHT})

\section{Conclusion}

This study presents a closed form expression for the diffuse field coherence between two arbitrary directional responses, expanded into a finite number of spherical harmonic coefficients. Due to the isotropy of the diffuse-field power, the coherence depends solely on the angular spectra of these directional responses, which can be either estimated, derived from a model, or measured. The effect of arbitrary spacings and orientations between sensors to the angular spectra can be conveniently expressed in the spherical harmonic domain. Finally, it is shown that the same principle can be applied without knowledge of the array characteristics, in the case that the overall array response is measured.


%

%

%

\ifCLASSOPTIONcaptionsoff
  \newpage
\fi



\bibliographystyle{IEEEtran}
\bibliography{APolitis_DiffCoherenceSensors}
%
%
%

%

%




\end{document}